# Virtualized 5G Air Interface Protocol Stack for Multi-Cell Coordination

*Óscar Carrrasco, Salva Díez, Jordi Calabug*

Sistelbanda SA, Valencia, Spain

*Abstract*—This article proposes a novel virtualized air interface protocol stack for next-generation wireless networks, that natively supports multiple radio access technologies and multi-point transmissions. Leveraging upon the concepts of softwarization of the air interface and virtualization, this design provides flexibility and scalability towards future advances in the radio access, whilst at the same time being backwards compatible with legacy technologies. This proposal enables the aggregation of multiple frequency bands and multiple technologies, without the need of modifying the operating procedures or the protocol of each supported technology. In the end, some challenges that have still to be addressed in future works are provided.

I. INTRODUCTION

Nowadays, the research activities of the mobile industry are focused on the definition of the fifth generation (5G) of mobile networks, which must address the challenges of a fully mobile and connected society consuming a very wide range of applications with vastly different characteristics and requirements [1-2]. The main challenges of future 5G systems are: the provision of a massive system capacity, due to the increase of the mobile data volume per area and the much larger number of devices compared with today; the support of very high data rates everywhere; the reduction of the end-to-end latency; and the reduction of the device cost and energy consumption [3].

Several technology concepts have been identified as potential enablers to be introduced into future 5G networks, which must fulfill key design requirements such as flexibility, scalability and agility from both the core network and the radio network perspective. Examples of these enablers are Network Function Virtualization (NFV) [4] and the Software Defined Networking (SDN) [5], which enable the separation of hardware from software, the softwarization of the network architecture, and a dynamic network topology reconfiguration. Additionally, the Software Defined Radio (SDR) paradigm [6], allows the virtualization of the air interface protocol stack.

A joint 5G research and innovation project, the 5G Infrastructure Public Private Partnership (5G-PPP), continues the initiated efforts of the Seventh Framework Programme (FP7) for research and development of the European Union. Within the research projects included in 5GPPP, the Speed-5G project [7] aims to achieve a significantly better exploitation of heterogeneous wireless technologies performing novel multi-cell coordination and exploiting new techniques for dynamically accessing the spectrum available in fragmented bands. The proposed techniques will take into account licensed, lightly licensed and non-licensed spectrum, as well as current and future radio access techniques.

Considering the challenges and requirements of future 5G systems, the different technology enablers already identified as well as the Speed-5G objectives, this article proposes a novel air interface protocol stack based on virtualization, running on radio access networks using the softwarization concept. This design provides the required flexibility in order to support multiple Radio Access Technologies (RATs), scalability depending on current and future needs; agility by adapting to different network requirements. The proposed air interface is capable of providing backwards compatibility with current radio technologies, such as Long Term Evolution (LTE) and the IEEE 802.11 (WiFi) standards, that will still play an important role in decades to come.

The rest of the paper is organized as follows. In Section II, the service types that were considered during the design of this protocol stack are briefly reviewed. Section III lists the general design requirements of a 5G air interface. In Section IV the proposal is described and its features and main advantages are listed in Section V. Finally, Section VI concludes this article by mentioning some of the challenges that should still be addressed in future works.

II. SUPPORTED SERVICE TYPES

The three main service types that were considered for the design of this novel air interface are a subset of the ones that are normally envisioned for 5G [8]. The service types under consideration are briefly reviewed here:

- Extreme Mobile Broadband (xMBB) requires extremely high data rate, low latencies and large coverage areas. The xMBB service type is shown in Fig.1(a). The main use cases for this service are, for instance, the future connected home and the virtual reality office. These use cases include scenarios that take place both indoor and outdoor, with various types of communicating devices, all requiring very good throughput experiences.

- Massive Machine-Type Communications (mMTC) is depicted in Fig.1(b). mMTC requires high scalability, wide coverage and deep indoor propagation. The main scope for this type of communication is to provide connectivity to up to tens of billions of connected devices, to enable the Internet of Things (IoT). A typical scenario representing this service type is the low-end IoT envisioned for demotics, smart metering and other non-critical applications.

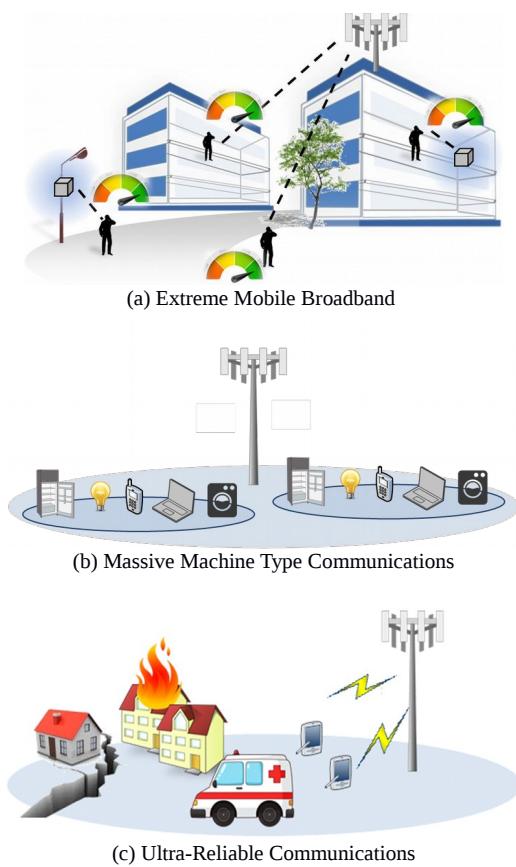

(a) Extreme Mobile Broadband

(b) Massive Machine Type Communications

(c) Ultra-Reliable Communications

Figure 1. The three main service types that were considered during the design of the proposed protocol stack [7].

- Ultra-Reliable Communications (URC) as shown in Fig.1(c) require reliable communications with very low latencies, that are needed to support safety critical applications, such as vehicle to anything (V2X) communications and other machine-type industrial applications. Typical scenarios for this service type include the health-monitoring sensors installed in an ambulance or the security sensors monitoring buildings and factories.

Despite the focus of the Speed-5G project is currently only on the three above scenarios, this proposal will be relevant and applicable for other scenarios as well, thanks to its flexibility and scalability.

## III. DESIGN REQUIREMENTS OF A 5G AIR INTERFACE

In this section, the key design requirements that a 5G air interface should be able to meet are reviewed.

### A. Extreme scalability

The air interface design should be able to scale efficiently in terms of provided data rate and number of served devices, while at the same time being able to provide reliability with fast access times and low latencies.

### B. Flexibility

The flexibility provided by SDN/NFV/SDR concepts allows splitting the network functionalities, being executed only where and when they are required within the mobile network. This capacity provides a break with traditional rigidity networks allowing the deployment of the layers and functions for cell coordination in an optimal way. Furthermore, the flexibility allows to group and ungroup a set of cells to perform specific joint coordination on demand. This is a great advantage since the procedures for cell coordination must include a variety of tasks, including the ones related to interference management or handover coordination, that are normally performed at higher layers of the protocol stack.

### C. Able to support multiple RATs

In the same way that a diverse range of scenarios and use cases is envisaged for 5G, a successful 5G air interface design should include the support of multiple technologies at different network layers, possibly very different from one another. Some examples of such disparate technologies that will play a role in next-generation wireless systems are, among others: network slicing, new multiple access schemes and waveforms, massive MIMO, millimeterwaves and mobile edge computing [9]. It is clear that the connectivity provided by LTE networks will play an important role in next-generation systems, although it is still not clear how this will be leveraged in 5G.

### D. Able to operate in a wide spectrum range

The air interface must be designed to work on a broad frequency range with very diverse characteristics, including different bandwidth and propagation proprieties. Lower frequencies will still remain essential to provide wide coverage areas and good building penetration characteristics. On the other hand, higher frequencies will be used mainly by small cells or macro cells combined with beamforming and advanced multiple antenna techniques. In any case, high frequencies will play an important role to support broadband users, as well as for backhaul data transport. Finally, the design of the air interface should take into account the growing importance of techniques to dynamically access the available spectrum [10], and that even more underutilized or unused frequency bands will be made available for wireless communications in the future [11].

### E. Being future-proof

Possibly one of the most important requirements of a 5G air interface is its capability to adapt to future changes and research advances the network access, as well as being able to support new services that are not yet imagined at present. This is one of the reasons to include SDN and NFV in this proposal, and to provide the support for SDR. In some way, this design principle subsumes the requirements that were listed so far.

### F. Being backwards-compatible

Current radio interfaces, such as LTE and WiFi will still play an important role in the decades to come. Because of this, it is therefore important that the air interface of next-generation networks still provides backwards compatibility with these technologies. For instance, the importance of LTE and its nearly ubiquitous connectivity should not be seen as a problem for 5G, rather a possibility to be exploited. The cardinal point is how to be able to integrate these technologies, whilst not hindering the flexibility of the air interface. It is also important that legacy technologies are supported without requiring changes in their operating procedures and protocols.

## IV. A Virtualized 5G Air Interface Protocol Stack

The proposed protocol stack is based on virtualization and software-defined components. In 5G, in fact, traditional networks will be replaced by logical networks deployed over slices [1]. A network slice represents a logical element that can be defined, activated or modified on demand. Even though a consensus on what the final goal of a slice has not yet been reached, it seems clear that this concept will play an important role in the architecture and design of next-generation wireless networks.

The novel proposal of a virtualized air interface protocol stack for 5G is shown in Fig.2. Together with the introduction of new layers, the key principles of this design are the virtualization of the components of the protocol stack, and the softwarization of the blocks above the physical layer.

In the rest of the section, the building blocks of the protocol stack are described in detail.

### A. 5G-RRM Layer

The Radio Resource Management (5G-RRM) layer implements strategies and algorithms for mobility, link management, resource and user allocation. Particularly, this layer is the main responsible for multi-cell coordination and the main enabler to achieve the maximization of spectral efficiency across different available frequency bands.

The RRM layer is the main signaling interface between multiple 5G cells and between the 5G cell and the User Equipment (UE). It is responsible for collecting the UE measurements and dynamically configuring the entire protocol stack of the air interface, depending on the current needs. It is also responsible for the management of the radio bearers and for controlling the Quality of Service (QoS) of the provided service.

### B. Data Forwarding Layer

Right below the 5G-RRM layer is the Data Forwarding layer which includes a Multipath TCP layer, an essential component for scheduling different TCP services over different radio bearers in parallel. These radio bearers are embedded in different data tunnels, in order to re-direct these services to one or more RATs of one or multiple cells. Together with the RRM layer, the Multipath TCP layer is the main enabler to support coordinated multipoint transmissions (CoMP) from multiple cells, as well as more advanced virtual multi-cell schemes.

### C. 5G-RRC

The 5G Radio Resource Control (5G-RRC) manages the control plane of the specific radio interfaces supported by one particular cell. It manages the different control procedures depending on the RAT. This layer can deploy entities per RAT, in order to specialize the control plane management for each technology. As an example, there is no RRC layer defined in WiFi, and therefore this will be configured to operate in transparent mode in this case.

In general, the services provided by this layer are the broadcasting of system information related to the non-access stratum, the broadcasting of system information related to the access stratum, the Paging procedures, the establishment, maintenance and release of a 5G-RRC connection between the UE and the RRM layer, various security functions including key management and establishment, the configuration and the operation of point to point radio bearers, functions related to mobility management and QoS management, and management operations related to UE measurement reports.

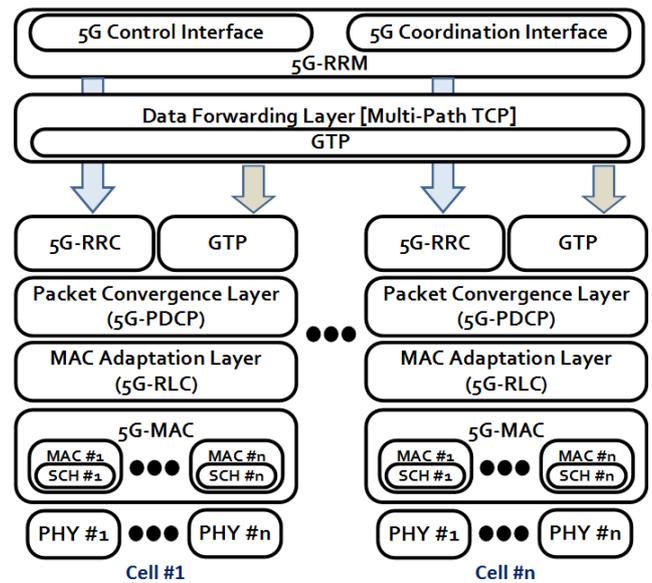

Figure 2. Speed-5G proposal of a 5G Virtualized Protocol Stack [7].

### D. 5G-PDCP

The main benefit provided by introducing a common Packet Data Convergence Protocol (5G-PDCP) layer for all RATs is that it enables one single security framework for all underlying technologies. The main functionalities provided by this layer are: ciphering and deciphering, header compression and decompression and in-sequence delivery of upper layer packet data units (PDUs).

### E. 5G-RLC (or 5G-Adaptation Layer)

The 5G Radio Link Control (5G-RLC) layer allows the management of multiple logical channels per UE. These logical channels may in turn be mapped to different underlying RATs. The 5G-RLC is also called the 5G-Adaptation layer as it is the main interface to lower stack layers responsible to access the channel. Together with the next layer described below, this is one of the key entities of this novel protocol stack.

### F. 5G-MAC Layer

Another key component of the proposal is the 5G Medium Access Control (5G-MAC) layer which is configured by the higher 5G-RRC. The 5G-MAC layer includes several different MAC entities that provide features that are RAT-dependent, such as the scheduling process for resource allocation of control and user data, and the Hybrid Automatic Repeat Request (HARQ) management, when required. Therefore, the proposed air interface includes the capability to have a specific scheduler per RAT. Each scheduler is univocally associated to one or more 5G-RLC logical channels, though one logical channel is not shared between schedulers.

The different scheduler entities are embedded in a common 5G-MAC structure that can access the different cell resources on demand, implementing as well a single data forwarding, signaling and configuration point towards upper and lower layers. This layer also provides common functions needed in order to support dynamic access operation on licensed, lightly-licensed and unlicensed frequency bands [12], that are shared among different RATs.

Due to the high flexibility of the propose interface, different radio technologies with different configurations can coexist within the same cell, enabling the support of multiple service types (e.g. xMBB and mMTC) simultaneously.

### G. PHY Layer

The physical (PHY) layer depends upon the underlying RAT. Some examples of technologies that will be supported by next-generation 5G systems are LTE and WiFi. Nevertheless, new technologies will play an important role in 5G, for example new waveforms, such as filter bank multi-carrier (FBMC) techniques [13].

The PHY layer is accessed by the higher 5G-MAC in a flexible and dynamic manner. Inside a cell, different RATs can coexist at the same time, and different carriers can be used by the same RAT but with different configurations. For instance, a FBMC access scheme can be configured for xMBB operation in one carrier of one precise licensed frequency band, and configured for mMTC in another unlicensed frequency band.

## V. BENEFITS INTRODUCED BY THE PROPOSED DESIGN

This section provides the key features and advantages of the proposed air interface.

The main characteristic of the protocol stack is that all layers may be virtualized, including the PHY procedures. They are software-controlled and can be dynamically configured on-the-fly depending on the current network needs and the specific scenario to be addressed. SDR is natively supported by our stack, as lower layers are transparent with respect to higher layers. Particularly, higher layers need not know the implementation details of lower layers, thanks to the introduction of the common 5G-MAC Adaptation layer and the 5G-MAC layer.

In addition to the flexibility and the scalability achieved through virtualization and software defined network paradigms, the split between the control plane (C-Plane) and the user plane (U-Plane) is another key point of our design. This split simplifies the 5G-RRM layer, which has the responsibility of coordinating all necessary functions, since the own network deployment enables the connectivity management between entities, the interface state management or the service provisioning.

Future network deployments may include ultra-dense network scenarios, which is seen as an important enabler to meet the demands of 5G [2]. This concept brings new challenges related to the way cells are coordinated and organized at the stack level. The possibility of managing ultra-dense networks is foreseen by this design and it is made possible thanks to the introduction of the 5G-RRM layer and the Data Forwarding layer, the two main responsible for coordinating multiple cells transmissions.

The proposed stack has to support real scenarios that include, for instance, heterogeneous networks composed only by pre-5G technologies. Even in this case, the architecture has to be able to manage and provide the required coordination for these cells. The proposed stack ensures backwards compatibility with respect to legacy technologies, such as LTE and WiFi. At the same time, backwards compatibility is not achieved at the cost of limiting the flexibility or the scalability of the overall protocol stack.

### A. Multi-cell Coordination

In order to achieve xMBB, coordination between cells is required. Certainly, 5G-RRM and 5G-MAC layers have the focus, within the protocol stack, of enabling the carrier aggregation over multi-cell transmissions. Fig. 3 shows an example of multi-cell coordination. It is important to stress that this is achieved in a completely transparent manner with respect to the end user, thanks to the split of the C-Plane and the U-Plane.

### B. Multi-RAT Cell Coordination

The support of multi-RAT is native to this stack design. Each cell can independently support specific RATs without hindering the operation of its neighbors or limiting the flexibility and agility of the overall network operation. Moreover, all building blocks of the stack need not be necessarily defined for each RAT at each cell. For example, in the case of WiFi, the 5G-RLC block will be in transparent mode, since no RLC is defined for this technology.

Coordination between multi-RAT cells is required in order to improve the total throughput and to ensure the end-to-end user QoS. The coordination is achieved at the 5G-RRM layer, which also implements the related QoS procedures.

## VI. CONCLUSIONS AND FUTURE WORK

This article has presented the virtualized protocol stack defined by the Speed-5G project for next-generation wireless networks, that meets the usual design requirements of a 5G air interface. Through the virtualization of the components of the protocol stack, this proposal provides native support for different radio technologies operating across different frequency bands. This design provides full flexibility towards future advances in access technologies, while at the same time being backwards compatible with legacy technologies such as LTE and WiFi.

Some of the challenges that are still to be addressed relate to the design of the optimal interfaces between the various layers of the stack, especially between the 5G-RRM and the 5G-PDCP, and between the 5G-RLC and the 5G-MAC layers. Additional challenges relate to the design of the signalization between the various layers of the stack, as well as taking into account how this is performed in case one or more of these entities is virtualized. Other challenges to be addressed in future works include the optimal selection and configuration of RATs operating across different frequency bands, and the definition of novel dynamic spectrum access techniques able to support the demands of 5G. Some of these challenges are currently being address by the Speed-5G project [7].

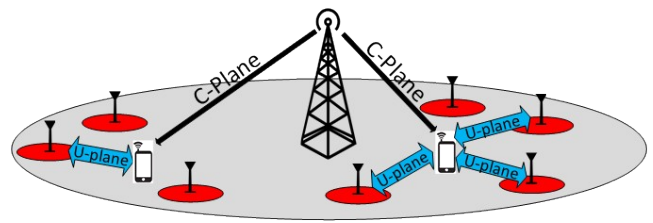

Figure 3. Example of the multi-cell coordination.

ACKNOWLEDGMENTS

The authors would like to thank the Speed-5G consortium for the useful discussion during the initial design of this protocol stack. This work has been funded in part by the European Commission H2020 programme under Grant Agreement N°: 671705.

REFERENCES

[1] NGMN Alliance, M. Iwamura, "NGMN View on 5G Architecture", IEEE Vehicular Technology Conference (VTC Spring), May 2015.

[2] J.F. Monserrat, G. Mange, V. Braun, H. Tullberg, G. Zimmermann, O. Bulakci, "METIS Research Advances towards the 5G Mobile and Wireless System Definition", EURASIP Journal on Wireless Communications and Networking, March 2015.

[3] P. Agyapong, M. Iwamura, D. Staehle, W. Kiess, A. Benjebbour, "Design considerations for a 5G network architecture", IEEE Communications Magazine, Vol.51, No.11, November 2014.

[4] ETSI ISG NFV (Operator Group), "Network Functions Virtualisation – Network Operator Perspectives on Industry Progress", Updated White Paper, October 2014.

[5] H. Kim and N. Feamster, "Improving network management with software defined networking," IEEE Communications Magazine, vol.51, no.2, February 2013.

[6] I. P. Belikaidis et al., "Context-aware Radio Resource Management Below 6 GHz for Enabling Dynamic Channel Assignment in the 5G era", EURASIP Journal on Wireless Communications and Networking (EURASIP JWCN), in preparation.

[7] U. Herzog et al., "Quality of service provision and capacity expansion through extended-DSA for 5G", Proc. of European Conference on Networks and Communications (EuCNC'16), June 2016.

[8] ICT-317669 METIS, Deliverable 1.5, "Updated scenarios, requirements and KPIs for 5G mobile and wireless system with recommendations for future investigations", April 2015.

[9] F. Boccardi, R.W. Heath, A. Lozano, T.L. Marzetta, P. Popovski, "Five disruptive technology directions for 5G", IEEE Communications Magazine, Vol.52, no.2, pp.74-80, February 2014.

[10] B. Jabbari, R.L. Pickholtz, M. Norton, "Dynamic spectrum access and management [Dynamic Spectrum Management]," IEEE Wireless Communications, Vol.17, No.4, pp.6-15, August 2010.

[11] S. Vassaki et al., "Interference and QoS Aware Channel Segregation for Heterogeneous Networks: A Preliminary Study", in Proc . of European Conference on Networks and Communications (EuCNC'16), June 2016.

[12] Oscar Carrasco et al., "Centralized Radio Resource Management for 5G small cells as LSA enabler," arxiv preprint: arXiv:1706.08057 (2017).

[13] A. Sahin, I. Guvenc, and H. Arslan, "A survey on multicarrier communications: Prototype filters, lattice structures, and implementation aspects", IEEE Communications Surveys & Tutorials, August 2014.